\documentstyle[epsf,prl,aps,multicol]{revtex}
\begin{document}
\title{Coupling between Edge and Bulk in Strong-Field Quantum Dots}
\draft
\author{S.-R. Eric Yang$^{a,b}$, and A.H. MacDonald$^{b}$}
\address{ $^a$ Department of Physics, Korea University, Seoul, Korea\\
 $^b$ Department of Physics, University of Texas at Austin, Austin, TX 78712, USA}
\date{\today}
\maketitle
\begin{abstract}

The maximum-density-droplet (MDD) state of quantum-dot electrons 
becomes unstable at strong magnetic fields to the addition of interior holes.  
Using exact diagonalization, we demonstrate that the first hole is located
at the center of the dot when the number of electrons $N$ is smaller
than $\sim 14$ and is located away from the center for larger dots.   The separation between
field strengths at which additional holes are introduced becomes small for large dots,  
explaining recent observations of a rapid increase in dot area when the magnetic field is
increased beyond the MDD stability limit.  We comment on correlations between interior
hole and collective edge fluctuations, and on the implications of these 
correlations for edge excitation models in bulk systems.  

\end{abstract}
\thispagestyle{empty}
\pacs{PACS numbers: 75.50.Pp, 75.10.-b, 75.30.Hx}
\begin{multicols}{2}

The fabrication of quantum dots has now reached an advanced state in which
the shape of the confinement potential and the number of electrons in a dot
can be tuned precisely \cite{review}.  In a strong magnetic field, quantum dots 
display a number of interesting many body effects \cite{pam,srey,st,jjp}, 
closely related to those exhibited by bulk two-dimensional (2D) electron systems in the 
quantum Hall regime.  Correlation effects become strong 
because the single-particle level spacing in this limit is  
small compared to the typical Coulomb interaction energy scale.
Some time ago we\cite{ahm} predicted that quantum dots 
in a strong magnetic field would form a {\em maximum density droplet} state with
approximately uniform electron density, close to that 
of a 2D system with a single filled Landau level ($\rho_0=(2\pi\ell^2)^{-1}$ where 
$\ell = (\hbar c/eB)^{1/2}$ is the magnetic length),
that would be stable over a wide field range. 
Indeed, features associated with the wide MDD state stability interval 
are prominent in {\em addition potential} 
{\it vs.} magnetic field traces measured \cite{ra,tho,ok} by using
transport resonances to track the gate voltages at which electrons are added to a quantum
dot.  The MDD state is a finite-size precursor and shares many attributes 
with bulk $\nu=1$ quantum Hall ferromagnet\cite{sondhi,jungwirth} states.
With increasing magnetic field, $\rho_0$ increases and Coulomb interactions eventually
make the MDD state unstable, favoring a state in which the average electron 
density is lowered.   Reimann {\it et al.} have concluded\cite{smr}, on the basis
of spin-density-functional calculations, that in large dots the density is first lowered by  
an edge reconstruction similar to those that occur in 
bulk systems at integer \cite{cdcc,ak} and fractional \cite{wanetal} filling factors.
In their self-consistent field calculations 
the MDD state evolves into a state with a modulation of 
charge density along the edge and a ring of electron density that break off from the
uniform density droplet.  For smaller dots, on the other hand, MacDonald {\it et al.} have
shown\cite{ahm} that the average density is lowered by introducing holes close to the center 
of the droplet.    

This Rapid Communication is motivated by recent experiments\cite{tho} of Oosterkamp {\it et al.},
who have measured {\em addition potential} {\it vs.} magnetic field traces for  
quantum dots containing $0$ to $40$ electrons.  These authors 
find that for $N$ larger than about $15$, the area occupied by the quantum dot electron cloud increases 
rapidly on the high-field side of the MDD stability region.  We have investigated 
the quantum dot phase diagram by performing exact diagonalization calculations,
assuming Zeeman fields large enough to ensure full spin polarization.
We find that holes are added to the interior of the dots with increasing field, 
increasing the electron cloud area; for $N>N_c$ ($N_c\approx 13-14$) the interior 
holes stay away from the center of the dot, while for $N<N_c$ the holes form a puddle at the center 
of the dot.  The phase diagram we obtain for hole number {\it vs.} field, is in 
qualitative agreement with experimental data.  We find that the rate at which
holes are added with field increases for larger dots, explaining the
observations of Oosterkamp {\it et al.}.  We also find that correlations
develop between the motion of holes in the interior of the dot and collective 
edge excitations of the dot, and discuss the nature of the corresponding 
coupling in bulk integer filling factor quantum Hall systems.    

We consider a system of 2D electrons confined by a parabolic potential, $V(r)=m^*\Omega^2r^2/2$.
and confine our attention here to the strong magnetic field limit, $\Omega/\omega_c \ll 1$. 
($\omega_c=eB/m^*c$, where $B$ is the magnetic field perpendicular to the 2D layer.)  In this limit
the symmetric gauge single-particle eigenstates in the lowest Landau level\cite{apology} 
are conveniently classified by an angular momentum index $m \geq 0$:
\begin{eqnarray}
\epsilon_m=\hbar\omega_c/2+\gamma(m+1), 
\end{eqnarray}
where $\gamma=m \Omega^2\ell^2$
We will drop the constant kinetic energy from subsequent discussion.
The Hamiltonian is invariant under spatial rotations about the axis
that is perpendicular to the 2D plane and passes through the center of the dot. 
It follows that the total angular momentum $M_z$ is a good quantum number. 
The Hamiltonian of the dot in the fully spin-polarized, lowest-Landau-level, Hilbert space is 
\begin{eqnarray}
&H&=\sum_m \gamma(m+1)c_{m}^{\dagger}c_{m}  \nonumber\\
&+&\frac{1}{2}\sum_{m_1',m_2',m_1,m_2}<m_1'm_2'|V|m_1m_2>c_{m_1'}^{\dagger}c_{m_2'}^{\dagger}c_{m_2}c_{m_1}.
\end{eqnarray}
The characteristic energy for the two-particle 
Coulomb matrix elements is $e^2/\epsilon \ell$, where $\epsilon$ is the host semiconductor
dielectric constant.  The eigenstates of this Hamiltonian depend only on the dimensionless
ratio between the two competing terms, $\tilde{\gamma}=\gamma/(e^2/\epsilon \ell)$.
The MDD state is the eigenstate of this Hamiltonian with the minimum angular
momentum allowed by the Pauli exclusion principle, $M_z \to M_{z}^{MDD} = N(N-1)/2$, and is the 
ground state of this Hamiltonian for strong confinement (large $\tilde \gamma$). 
We find it useful below to denote occupation number eigenstates by lists 
ordered by increasing angular momentum.  For example, in this notation the $N=4$ MDD state is 
$(\bullet,\bullet,\bullet,\bullet,\circ,\circ,...)$,
where the symbols $\bullet$ and $\circ$ denote occupied and unoccupied single-particle states.

\begin{figure}
\center
\centerline{\epsfysize=2.3in
\epsfbox{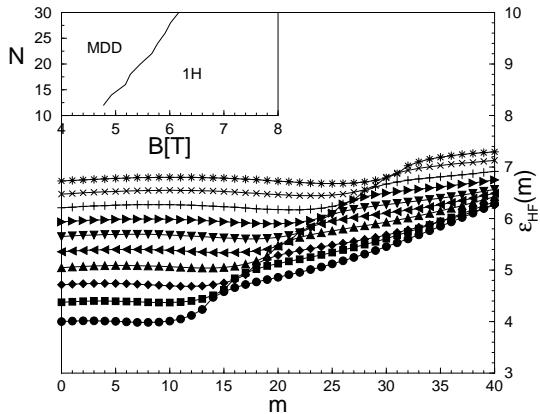}}
\begin{minipage}[t]{8.1cm}
\caption{Hartree-Fock quasi-particle energies $\epsilon_{HF}(m)$ of the MDD state 
as a function of $m$ for different values of $N$.  
$\tilde{\gamma}$ is chosen, for each $N$, so that the MDD is 
close to its instability point; $\tilde{\gamma}=$0.077, 0.08, 0.082, 0.085, 0.0875, 0.0925,
0.0975, 0.1, 0.108, 0.113 for $N=30, 28, 26,...  , 14, 12$, respectively.  
The inset displays the HF phase boundary between the MDD
state and one-hole states for a dot with $\hbar\Omega=3 {\rm meV}$.    
The dimensionless confinement strength parameter $\tilde \gamma$, the field in Tesla $B$,
and the confinement oscillator frequency in meV are related by 
 $B [{\rm Tesla}] = [0.131(\hbar\Omega[{\rm meV}])^2/\tilde \gamma]^{2/3}$.
}
\label{figure1}
\end{minipage}
\end{figure}
In Fig. 1 we have plotted Hartree-Fock quasiparticle energies 
,$\epsilon_{HF}(m)$, for $m=0$ to $m=40$ for 
MDD states with $N=12,14,...,28,30$.  
($\epsilon_{HF}(m)=\gamma( m+1)+\sum_{m'=0}^{N-1}U_{m,m'}$ where 
$ U_{m,m'}=<mm'|V|mm'>-<mm'|V|m'm>$)  
The confinement strengths used in these plots are, for each $N$, near the MDD stability limit. 
In HF theory, the MDD becomes unstable when an unoccupied interior quasiparticle energy 
exceeds the edge quasiparticle energy, {\it i.e.} when $\epsilon_{HF}(m^*) > \epsilon_{HF}(N-1)$
for $m^* < N-1$.  We see from Fig. 1 that holes are first introduced near the 
center of the dot $ N \le 12$ and gradually move outward as $N$ increases.  This qualitative prediction
of Hartree-Fock theory is generally confirmed by the exact diagonalization calculations we 
discuss below.  

To understand the role played by correlations we first discuss a small dot
for which an accurate analytic calculation is possible.  
For $N=3$ the MDD has $M_z^{MDD}=3$ and HF theory predicts that when $\gamma$ is reduced
the first hole is introduced in the $m^*=0$ state, increasing the $N=3$ ground state 
angular momentum to $M_z=6$.
The exact single hole ground state is obtained by diagonalizing the many-particle Hamiltonian
in the space of states with $M_z=6$.  Fig. 1 suggests that the probability of occupying 
single-particle states with large $m$ is small, since quasiparticle energies increase 
rapidly outside the MDD.  If we can neglect the possibility of occupying states with $m > m_c =4$,
the only additional state in the Hilbert space with $M_z=6$ has the interior hole at 
$m=1$ rather than $m=0$ and excites an edge magnetoplasmon at the edge\cite{ahm,zulicke} 
by making a particle-hole excitation from $m=3$ to $m=4$.  The many-particle Hamiltonian matrix 
in this two-dimensional Hilbert space is
\begin{equation}
\left(
\begin{array}{cc}
 9  \gamma + U_{1,2}+U_{1,3}+ U_{2,3} & U_{0413} \\U_{0413}
& 9 \gamma + U_{0,2}+U_{0,4}+ U_{2,4} 
\end{array}\right),
\end{equation}
where
\begin{eqnarray}
U_{m_1m_2m_3m_4}&=&<m_1m_2|V|m_3m_4>\nonumber\\
&-&<m_1m_2|V|m_4m_3>.
\end{eqnarray}
The interaction contributions to the matrix elements in Eq. 3 in $e^2/\epsilon\ell$ units
are $0.940$ and $1.114$ respectively along the diagonal and $0.083$ in the off-diagonal 
term.  The off diagonal term introduces correlation between hole motion and collective
edge excitations in this small dot, reducing the ground state energy to 
$ 9 \gamma + 0.907 e^2/\epsilon \ell$.  The $N=3$ ground state energy, in this approximation,
is $9 \gamma + 0.907 e^2/\epsilon \ell$, compared to the MDD state energy which is 
$6 \gamma + 1.204 e^2/\epsilon \ell$.  Note that correlations between bulk and edge 
shift the MDD stability limit from $\tilde \gamma = 0.088$ in the Hartree-Fock approximation 
to $\tilde \gamma = 0.099$.  An exact diagonalization for $N=3$ confirms the accuracy of this
Hilbert space truncation.  

\begin{figure}
\center
\centerline{\epsfysize=2.3in
\epsfbox{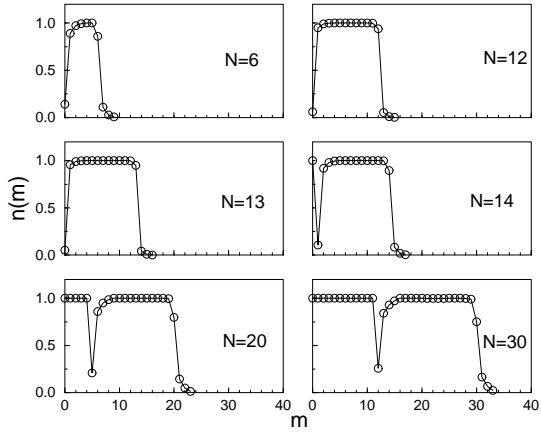}}
\begin{minipage}[t]{8.1cm}
\caption{
Occupation number distributions for the one-hole MDD states 
of $N=6,12,13,14,20,30$ quantum dots.
}
\label{figure2}
\end{minipage}
\end{figure}
Fig. 2 summarizes our numerical exact diagonalization results for one hole states
at larger $N$ by plotting mean-occupation numbers for all single-particle states.
For $N=6$, the HF approximation places the hole at $m^*=0$, yielding a total $M_z=21$. 
The exact single-hole state occurs at the same value of $M_z$, but the hole position and 
the dot edge have correlated quantum fluctuations, evidenced in Fig. 2 by the finite probability
for finding the hole in the $m=1$ state.  Equal time correlation functions, not 
illustrated here, demonstrate that 
the hole has a higher probability of being on the same side of the system as edge density
modulations.  For $N=12$, we again find that the exact single-hole state has the 
same $M_z$ as the Hartree-Fock state with $m^*=0$, $M_z=78$, and correlated bulk-hole
edge excitations.  This pattern continues for larger particle numbers with the nominal hole
angular momenta moving out at larger $N$.  For $N=30$, for example, the nominal angular 
momentum of the single-hole is $m^*=12$.  Note that the hole angular momentum has 
fluctuations only to the high-angular momentum side of its nominal value.  This property
is a combined consequence of angular momentum conservation and the chiral\cite{wenedge}
nature of the collective edge excitations with which it is correlated.   
The one-hole nominal locations for $N=6,12,13,14,20,30$ 
$m^*= 0,0,0,1,5,12$ respectively.  The total hole occupation number determined by 
summing the depleted occupation numbers inside the quantum dots in Fig. 2 is accurately quantized 
at 1, demonstrating that holes and edge excitations are correlated but that charge
fluctuations from bulk to edge or {\it vice versa} is negligible. 

\begin{figure}
\center
\centerline{\epsfysize=2.3in
\epsfbox{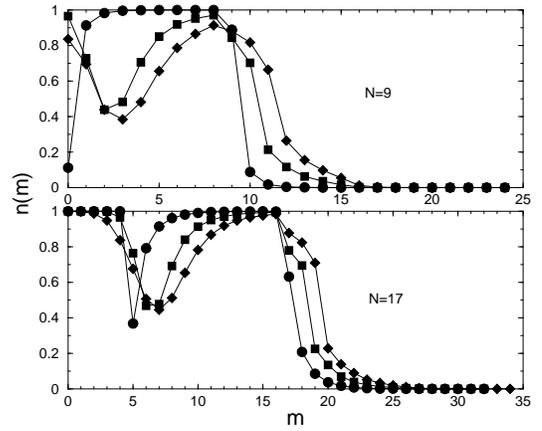}}
\begin{minipage}[t]{8.1cm}
\caption{Occupation number distributions for 
one-hole (circle), two-hole (square), and three-hole (diamond) MDD states 
for $N=9$ and $N=17$.
}
\label{figure3}
\end{minipage}
\end{figure}

For each particle number $N$, the number of holes in the ground state increases 
as $\tilde \gamma$ decreases, with the number of bulk holes accurately conserved
to larger hole numbers for larger $N$.  In Fig. 3, we plot occupation number distributions
for the one-hole two-hole and three-hole states at $N=9$ and $N=17$.  We find that the
average number of interior holes in the nominal two and three hole states deviates from integer
values by 
$0.007$ and $0.05$ for $N=9$ and $0.003$ and $0.006$ for $N=17$, demonstrating accurate 
interior hole-number conservation.  The one-hole and two-hole $M_z$ values are $45$ and $50$ for 
$N=9$ and $148$ and $158$ for $N=17$, giving nominal first-hole and second-hole
$m^*$ values of $0$ and $4$ for $N=9$ and $5$ and $7$ for $N=17$.  It is clear from
Fig. 3, however, that correlations within the two-hole system make these assignments,
based on the sequence of ground state angular momenta values, less meaningful.  For States with
two or more holes, correlations among the holes and correlations between the holes and 
the edge system are both important. 
\begin{figure}
\center
\centerline{\epsfysize=2.3in
\epsfbox{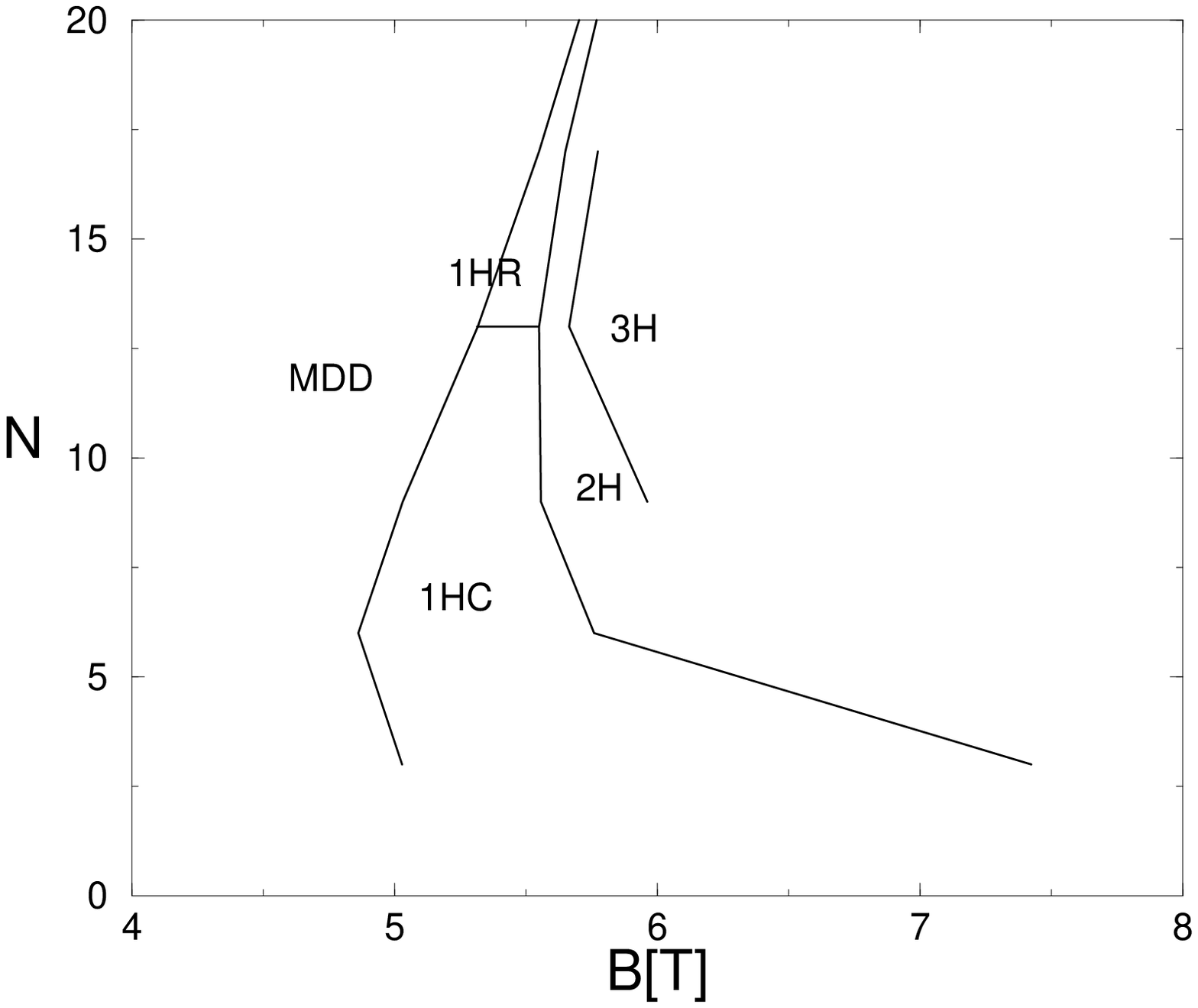}}
\begin{minipage}[t]{8.1cm}
\caption{Phase diagram near the stability limit of the MDD state.
The symbol MDD stands for the maximum density droplet, while
1HC and 1HR represent represent respectively centered ($m^*=0$) and 
ring ($m^*>0$) single-hole states. 2H and 3H represent states with two and three holes.
States with more than three holes have not been studied but will occur at 
larger fields.  Classifying states near the MDD limit by the number of holes 
is meaningful when charge fluctuations between the interior of the MDD and its edge 
are negligible.
}
\label{figure4}
\end{minipage}
\end{figure}

To facilitate comparison with experiment, we have plotted a phase diagram of 
MDD related states in Fig.4 {\it vs.} particle number $N$ and field, parameters that 
are known experimentally, using a typical value for the parabolic confinement 
strength $\hbar\Omega=3meV$.  We have not considered states
with more than three holes because of the growing difficulty of carrying out accurate 
exact diagonalization calculations, and expect that the region at the top right of this
phase diagram should be occupied by states with four and more interior holes.
Note that the spacing in field between hole number addition points becomes small for
large dots, explaining the sudden increase in quantum dot area observed by
Oosterkamp {\it et al.}.  The stability regime of multiple hole states is 
seriously underestimated by Hartree-Fock and density-functional-theory calculations
which are unable to account for the strong correlations possible in the quantum
Hall regime. 

The sign of cusps in the magnetic field dependence of
the addition spectrum contains valuable information about the shape of the phase boundaries. 
Oosterkamp et al \cite{tho} measured is the magnetic field dependence of
the addition spectrum given by $\mu_N=E_N-E_{N-1}+\hbar\omega_c/2$, where 
$E_N=\gamma(N+M_z)+U(N,M_z)$ is the groundstate energy of an $N$-electron system. (Here $U(N,M_z)$
is the total interaction energy).  As the applied magnetic field increases,  
groundstate level crossings in the $N-1$ and $N$ particle systems lead to positive and
negative cusps\cite{srey}, respectively.  
For $N>8$ the  phase boundary between the 
MDD and one-hole states is an increasing function of magnetic field (see Fig.4).
This implies that the $N-1$-electron dot will cross the phase boundary before the $N$-electron
dot does.  The measured upward cusps are consistent with this expectation.
Upward cusps are also seen between one-hole and two-hole states.

The strong correlations that we find between interior holes and the edge excitations of 
quantum dots also applies to bulk quantum Hall systems. 
These correlations are important because of the long-range of the Coulomb interaction,
which causes charge fluctuations in the two subsystems to interact strongly.
Recent scanning probe studies of 2D electron systems in the quantum
Hall regime\cite{2Dscanprobe} have made it clear that real experimental samples always 
have low-energy interior hole and electron low-energy degrees of freedom.
At low temperatures charge carriers in the bulk are localized so these disorder
induced degrees of freedom do not influence the quantum Hall effect.
They may, however, influence other properties of the chiral edge state system.
Collective excitations at the edge, will induce electric potentials in the interior
that will excite the localized hole system.  Similar effects cause semiclassical edge
magnetoplasmon\cite{mikhailov} excitations to be influenced by the bulk conductivity.
The possible importance, at experimental temperature and voltage scales, 
of these interactions for the tunneling density-of-states\cite{kanefisher} and 
other properties measured at the edge deserves more careful consideration. 

We are indebted to D.G. Austing and S. Tarucha for 
valuable discussions.  This work was supported by the National Science 
Foundation under grant DMR0115947, by the Welch foundation, and by
the Korea Research Foundation under Grant KRF-2000-015-DP0125.

\end{multicols}

\end{document}